\documentclass[conference]{IEEEtran}
\IEEEoverridecommandlockouts

\usepackage{cite}
\usepackage{amsmath,amssymb,amsfonts}
\usepackage{algorithmic}
\usepackage{graphicx}
\usepackage{textcomp}
\usepackage{xcolor}
\usepackage{url}
\usepackage{hyperref}
\usepackage{microtype}
\usepackage{booktabs}
\usepackage{multirow}
\usepackage{array}
\usepackage{colortbl}
\usepackage{pifont}
\usepackage[most]{tcolorbox}
\usepackage[english,bidi=default]{babel}

\newcommand{\testasvi}{\textbf{LA19}}
\newcommand{\testasvii}{\textbf{LA21}}
\newcommand{\testitw}{\textbf{ITW}}
\newcommand{\testml}{\textbf{ML}}

\begin{document}


\title{Textual Acoustic Grounding for Generalizable LLM-Based Deepfake Voice Detection}



\author{
\IEEEauthorblockN{
\begin{minipage}{\textwidth}
\centering
Yassine El Kheir\IEEEauthorrefmark{1}\textsuperscript{,}\IEEEauthorrefmark{3}, 
Xin Wang\IEEEauthorrefmark{2}, 
Wanqing Ge\IEEEauthorrefmark{2}, 
Tim Polzehl\IEEEauthorrefmark{3}, 
Sebastian Möller\IEEEauthorrefmark{1}, 
Junichi Yamagishi\IEEEauthorrefmark{2}
\end{minipage}
}\vspace{-0.4cm}
\and
\IEEEauthorblockA{
\begin{minipage}{\textwidth}
\centering
\IEEEauthorrefmark{1}\textit{German Research Center for Artificial Intelligence (DFKI)}, Berlin, Germany \\
\IEEEauthorrefmark{2}\textit{National Institute of Informatics}, Tokyo, Japan \\
\IEEEauthorrefmark{3}\textit{Technical University of Berlin}, Berlin, Germany \\[0.05cm]
yael02@dfki.de
\end{minipage}
}
}

\maketitle
\vspace{-1cm}


\begin{abstract}
Deepfake voice detection suffers from poor generalization across unseen domains. While Audio Large Language Models (ALLMs) show promise, the modality gap between continuous audio embeddings which capture the subtle acoustic details necessary for deepfake detection and the semantic space of LLMs remains a critical, underexplored bottleneck. We address this by benchmarking diverse audio encoders integrated with Qwen LLMs (0.5B to 7B parameters). First, we demonstrate that fine-tuning the LLM alone risks out-of-domain overfitting, making a frozen LLM a stronger, resource-efficient baseline. Second, to explicitly bridge the modality gap, we introduce a cross-modal prompting strategy that injects linguistic-knowledge-driven acoustic features (via openSMILE) as structured text tokens. This explicit textual grounding not only enhances the frozen baseline but also makes LLM fine-tuning more effective.
Ultimately, our approach demonstrates state-of-the-art resilience on the out-of-domain ITW and MLAAD benchmarks, yielding over \textbf{16.2\%} absolute improvement in Macro-F1 over existing ALLM baselines while maintaining competitive in-domain performance. All models reported in this work are \href{https://huggingface.co/01Yassine/AudioLLM-Deepfake-Detection}{publicly available}.
\end{abstract}


\section{Introduction}
\label{sec:intro}

Cloning a target speaker's voice has been made easier by recent deep-learning-based speech synthesis technologies~\cite{chenNeural2025, tanNeural2023}.
Concurrently, the ease of forging a victim's voice, resulting in deepfakes, has garnered significant attention from both the research community and society at large~\cite{liSurvey2025}.

A countermeasure is to use machine learning techniques and build automatic detectors that discriminate deepfake voices from real human voices.
This topic has been investigated by the research community for the last decade, resulting in many architectures~\cite{jung2022aasist, tak2021rawnet2}, feature pipelines~\cite{tak2022wav2vec, yang2019lfcc}, databases~\cite{muller2024mlaad, todisco2019asvspoof}, and challenge benchmarks~\cite{todisco2019asvspoof, liu2023asvspoof2021, wang2024asvspoof5}.
Despite the progress, a key issue stands out: it remains challenging to build a robust detector that learns from a fixed set of training data and generalizes to unseen deepfake types~\cite{liu2023asvspoof2021, muller2022ood, wang2024asvspoof5}.

Large Language Models (LLMs), including multimodal Audio LLMs (ALLMs), brings new possibilities to this domain. By leveraging emergent reasoning capabilities, broad knowledge of general audio and world knowledge encoded within massive text-audio pretrained networks, ALLMs hold the potential to evaluate acoustic anomalies more robustly than traditional classifiers~\cite{gu2025allm4add, li2025dfallm, xie2026ftgrpo}.


However, applying LLMs to deepfake forensics introduces a critical cross-modal alignment challenge: how to effectively map audio representations into the semantic space that LLMs understand natively. Existing studies have largely treated this as a brute-force parameter optimization problem, focusing either on zero-shot inference~\cite{xuHoliAntiSpoof2026} or full parameter fine-tuning~\cite{gu2025allm4add, xie2026ftgrpo}, leaving the configuration of the audio encoder unexplored.

In standalone deepfake detection systems, self-supervised models like Wav2Vec2 are known to be highly effective audio encoders~\cite{kheirDeepFense2026}. However, it remains an open question whether they are equally effective in LLM-based architectures. Alternatively, because LLMs operate inherently in a semantic space, an audio encoder like Whisper, which maps acoustic representations to higher-level linguistic knowledge~\cite{ogunremiTranscribe2025}, might be more suitable for bridging this modality gap. 
Furthermore, because LLMs are heavily biased toward semantic understanding, they often fail to capture the acoustic cues indicative of deepfakes when relying solely on audio tokens.

We hypothesize that explicitly anchoring the audio embeddings with structured natural language descriptors can bridge this modality gap. By providing the LLM with expert linguistic knowledge via text prompts, we can ground its classification in explicit evidence. We are therefore motivated to address the following research questions: 


\textbf{RQ1) The Efficiency Baseline:} What is the optimal audio encoder integration strategy to maximize out-of-domain robustness, and can we achieve this without the computational burden of fine-tuning the LLM backbone?

\textbf{RQ2) The Cross-Modal Bridge:} How does 
prompting using explicit textual description of acoustic cues 
enhance this baseline, and how does this textual bridge interact with the LLM adaptation strategies explored in RQ1?

Together, RQ1 and RQ2 trace a path from an efficient frozen baseline to a fully cross-modal grounded detector. Our contributions to address these questions are:
\begin{itemize}

  \item We systematically compare audio encoder families; Whisper~\cite{radford2023whisper}, Wav2Vec2~\cite{baevski2020wav2vec}, EAT~\cite{eat}, and the encoder modules of speech codecs; integrated with Qwen LLMs (0.5B--7B) under varied fine-tuning strategies.

  \item We introduce features extracted using the knowledge-driven \textbf{openSMILE} framework~\cite{eyben2010openSMILE}.
  These features cover a wide range of acoustic properties of the speech utterance and are fed to the LLM in a structured text format.
\end{itemize}
To the best of the authors' knowledge, both research questions are not covered by existing literature. 

A notable finding is that fine-tuning the audio encoder is the dominant factor for improving generalization. 
In contrast to prevailing assumptions, we find that fine-tuning the LLM alone 
degrades out-of-domain performance,
and a frozen instruct-LLM is a more robust baseline~\cite{gu2025allm4add, li2025dfallm, xuHoliAntiSpoof2026, xie2026ftgrpo, guo2026sddapallm}. However, the degradation is mitigated 
when the LLM is explicitly prompted with textual descriptions of the acoustic features (extracted via openSMILE); LoRA-based LLM adaptation becomes highly synergistic and yields the best overall performance.

In the rest of the paper, related work is presented in \S\ref{sec:related}. The LLM-based framework, audio encoders, openSMILE features, and LLM fine-tuning strategies are described in \S\ref{sec:method}.
Experiments on the two research questions are presented in \S \ref{sec:exp:rq1} and \S\ref{sec:exp:rq2}, respectively. Comparison with the state of the art systems are presented in \S\ref{sec:compare}. 
Conclusion is drawn in \S \ref{sec:conclusion}.

\section{Related Work}
\label{sec:related}

With model components pre-trained on large-scale data, the ALLM may generalize better to various types of deepfake. ALLM4ADD~\cite{gu2025allm4add} was the first work to formulate deepfake voice detection as a question answering task and fine-tune ALLMs for the task.
However, this study used only pre-trained ALLMs without exploring audio encoders design.
DFALLM~\cite{li2025dfallm} examined the role of the audio encoder but only compared Whisper with Wav2Vec2-BERT~\cite{chung_w2v-bert_2021}, both of which were designed for the automatic speech recognition (ASR).

FT-GRPO~\cite{xie2026ftgrpo} introduces fine-tuning based on reinforcement learning (RL) and structured frequency-time rationales.
SDD-APALLM~\cite{guo2026sddapallm} reformulates the problem through the lens of \emph{acoustic evidence accessibility}, showing that explicitly providing time-frequency representations guides the LLM toward acoustic cues useful for deepfake detection.
This work is related to ours in terms of investigating the impact of handcrafted audio encoder features.

Our study addresses research questions not covered by existing work.
First, we conduct a systematic comparison of different types of audio encoders, including Whisper, Wav2Vec2, EAT, and two types of speech codec encoders.
Speech codecs, which compress speech utterances into sequences of discrete tokens, are important components in ALLM-based speech generation~\cite{chenNeural2025} but have not been investigated for deepfake voice detection.
Furthermore, the synergy between the choice of audio encoders and the training strategy of the LLM backbone remains unexplored.

Another novelty is the use of text descriptions of acoustic features extracted using the openSMILE framework.
Compared with the time-frequency representation used in SDD-APALLM~\cite{guo2026sddapallm}, our openSMILE-based features cover acoustic cues beyond time-frequency representations, such as the intonation of the input utterance. While openSMILE features have recently been applied to non-DNN-based detectors~\cite{pascu_easy_2025}, leveraging these descriptive acoustic prompts to help LLMs detect synthetic speech artifacts represents a novel approach.


\section{Proposed System}
\label{sec:method}

\subsection{Architecture}

\begin{figure}[!t]
    \centering
    \includegraphics[width=0.95\linewidth, trim=5 200 500 5,clip]{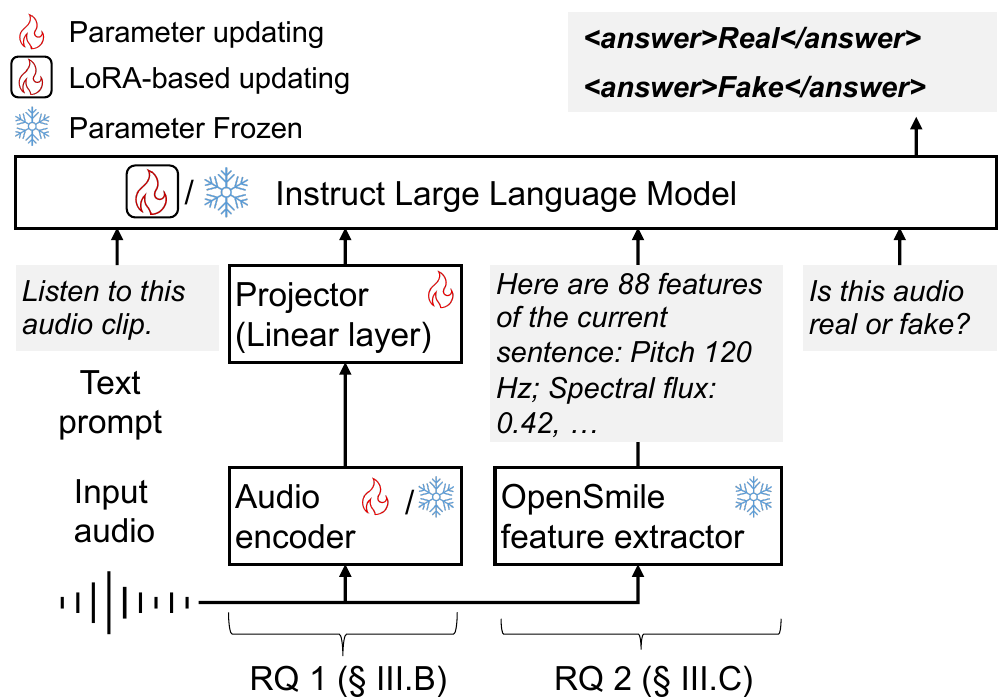}
    \caption{Overall illustration of proposed LLM-based deepfake voice detector.}
    \label{fig:overall}
\end{figure}

Our base framework, illustrated in Fig.~\ref{fig:overall}, follows the standard ALLM structure~\cite{gu2025allm4add, li2025dfallm, xuHoliAntiSpoof2026, xie2026ftgrpo, guo2026sddapallm} and includes an audio encoder $\mathcal{F}_{\mathrm{enc}}$, a lightweight linear projector $\mathcal{P}$, and an LLM backbone $\mathcal{F}$.
Given an input utterance $x$, a sequence of audio embedding vectors is extracted:
$\mathbf{T}_{\mathrm{aud}} = \mathcal{P}(\mathcal{F}_{\mathrm{enc}}(x)) \in \mathbb{R}^{T \times D}$,
where $T$ and $D$ are the number of tokens and the dimension of each vector, respectively. Note that the linear projector $\mathcal{P}$ is always fine-tuned alongside the audio encoder, even when the LLM backbone remains frozen.
The audio embeddings are concatenated with the embeddings of the text prompt, and the LLM backbone produces the real/fake output in text format, surrounded by \texttt{<answer>} tags.

Based on the framework, we systematically investigate two technical aspects: the choice of audio encoder and its integration with the LLM (\S\ref{sec:method:encoder}), and whether knowledge-driven acoustic features improve the detection (\S\ref{sec:method:openSMILE}).

\subsection{Audio Encoder and LLM Configuration}
\label{sec:method:encoder}

\subsubsection{Choice of Audio Encoder and Fine-Tuning Strategy}

Existing ALLM-based deepfake voice detection systems predominantly use Whisper~\cite{radford2023whisper} to process audio.
However, since Whisper was trained for speech recognition and tends to discard non-linguistic acoustic details, features potentially relevant to deepfake voice detection may be neglected~\cite{li2025dfallm}.

Many audio processing modules have been applied to non-LLM-based deepfake voice detectors, but their integration with LLMs has received limited attention.
One family is pre-trained SSL models such as Wav2Vec2~\cite{baevski2020wav2vec}.
Hidden representations from intermediate layers of SSL models have been found to be effective for deepfake voice detection~\cite{kheir2025, xiaoLayerWise2025}.

In addition, deep-learning-based neural speech codecs are applicable as audio encoders~\cite{kumar2023dac}.
They decompose speech utterances into sequences of quantized vectors and reconstruct utterances with high fidelity.
Whether these quantized features remain useful for deepfake voice detection is an open question.

Hence, we investigate the following audio encoders:
\begin{itemize}
    \item \textbf{Whisper encoder:} The default audio encoder in many ALLMs, pretrained on large-scale ASR data.
    
    \item \textbf{Wav2Vec2:} A pretrained model via contrastive self-supervised learning over raw speech waveforms at 50\,Hz. Its training objective encourages preserving acoustic features at different linguistic levels, making it a strong candidate for deepfake voice detection.

    \item \textbf{Environmental Audio Transformer (EAT)~\cite{eat}:} A Transformer-like model pretrained on large-scale environmental sound data. We include it as a representative non-speech SSL model.

    \item \textbf{Neural Audio Codecs:} Speech codecs, such as Descript Audio Codec (DAC)~\cite{kumar2023dac} and SpeechTokenizer~\cite{zhang2023speechtokenizer}, which use convolutional layers and vector quantization to encode speech into sequences of discrete vectors. 
    We extract the continuous feature embeddings from the first $N$ residual vector quantization (RQV) codebooks (denoted as cont:N). Specifically, we set cont:4 for SpeechTokenizer and cont:6 for DAC. 

\end{itemize}

For each audio encoder, we consider three integration strategies:
1) \emph{Freezing} (FR): no updating of the pre-trained audio encoder; 2) \emph{Fine-tuning} (FT): updating all parameters given gradients from the LLM backbone; 3) \emph{Weighted aggregation} (WA): no parameter updates, but trainable weights compute a weighted sum of outputs across all Transformer blocks inside the encoder.
The third strategy is widely used for deepfake voice detection~\cite{kheir2025, xiaoLayerWise2025} but not applicable to the speech codecs that do not use the Transformer architecture.

\subsubsection{Choice of LLM Backbone and Fine-Tuning Strategy}
\label{sec:method:llm}

We use open-source LLM models from the Qwen2.5-Instruct family~\cite{qwen2} at three scales: 0.5B, 3B, and 7B.
One option is to freeze its pretrained parameters.
Because the Qwen2.5-Instruct models have been instruction-tuned, given appropriate prompts, they can generate binary decisions (real/fake) without fine-tuning. Another option is to fine-tune the LLM backbone using LoRA~\cite{hu2022LoRA}, as done in many existing studies.
The fine-tuning is accomplished given a ground-truth text target \texttt{<answer>real</answer>} (or \texttt{fake}) alongside the text prompt and audio input.

We additionally evaluate Qwen2.5-Omni-3B~\cite{qwenomni}, a native end-to-end 
multimodal LLM, with full LoRA adaptation following an existing recipe~\cite{xie2026ftgrpo}.


\subsection{openSMILE Text-Token Injection}
\label{sec:method:openSMILE}

LLMs are predominantly trained on text corpora and may not 
focus on the acoustic details encoded in audio encoder's outputs.
Furthermore, the audio encoder's outputs consist of hidden representations from DNNs, 
which lack explicit interpretability.
To provide the LLM with explicit acoustic features, we extract a compact, expert-validated set of features using openSMILE eGeMAPSv2~\cite{eyben2010openSMILE}. We select eGeMAPSv2 specifically for its optimal balance between low dimensionality (88 features) and high interpretive relevance to voice physiology. 

The openSMILE features are originally represented as a continuous numerical vector. To make them compatible with the LLM and maximize interpretability, we serialize them into a descriptive text string. Rather than relying solely on raw acronyms, we explicitly expand the feature names into text description
(e.g., ``Average Fundamental Frequency'' instead of just F0, or ``Average First Formant Frequency'' for F1). The resulting serialized text block adds only approximately $1500$ tokens to the prompt. 
An example openSMILE prompt is shown below. 
It is inserted after the continuous acoustic embeddings extracted by the audio encoder (Figure~\ref{fig:overall}).




\begin{tcolorbox}[
    colback=blue!5,
    colframe=blue!75!black,
    title={openSMILE Prompt Structure},
    fonttitle=\small\bfseries,
    fontupper=\small\ttfamily,
    arc=1mm,
    boxrule=0.5pt
]
Here are 88 speech features of the current speaker's sentence. The features are: Average Fundamental Frequency in Semitones from 27.5 Hz: 33.758, Standard Deviation of F0 (normalized): 0.101, [...] Equivalent Sound Level (dBp): -13.031.
\end{tcolorbox}

The text-based openSMILE prompt provides \emph{directly accessible} evidence that the LLM can compare, contrast, and evaluate using its encoded knowledge (e.g., typical human pitch ranges, natural harmonic-to-noise ratios).
This is conceptually aligned with the acoustic-evidence accessibility principle of~\cite{guo2026sddapallm}, but operates via text rather than visual tokens.
Integrating the text-based features requires no architectural change to the LLM backbone. 
This token overhead is negligible for modern LLMs with large context windows and adds minimal latency. 

\section{Experiment Setup}
\label{sec:exp:setup}

We conduct two experiments to address the two research questions, both using the setup described below.

\noindent\textbf{Audio Encoders:} We implement our pipeline using five encoder families: 300M-parameter Wav2Vec2.0 XLS-R via Fairseq\footnote{{https://github.com/facebookresearch/fairseq/blob/main/examples/wav2vec/xlsr}}, Whisper\footnote{\url{https://huggingface.co/openai/whisper-large-v3}}, EAT (EAT)\footnote{\url{https://huggingface.co/worstchan/EAT-base_epoch30_pretrain}}, SpeechTokenizer (\texttt{hubert\_avg} config)\footnote{\url{https://github.com/zhangxinfd/speechtokenizer}}, and DAC(16\,kHz)\footnote{\url{https://github.com/descriptinc/descript-audio-codec}}. For the neural codecs, continuous feature embeddings from the first 4 and 6 RVQ codebooks are extracted from SpeechTokenizer and DAC, respectively.


\noindent\textbf{Training Data:}
We follow the standard protocol and use the training partition of the \textbf{ASVspoof 2019 (LA)} dataset~\cite{todisco2019asvspoof}. This partition contains $25{,}380$ utterances ($2{,}580$ bona fide and $22{,}800$ spoofed) from $20$ speakers. Spoofed speech is generated using six different text-to-speech (TTS) and voice conversion algorithms. We process the {full, uncropped utterances} during both training and evaluation.

\noindent\textbf{Evaluation Benchmarks:} To rigorously test generalization, we evaluate on four distinct benchmarks: \textbf{1) test set of ASVspoof 2019 LA (\testasvi)}, our in-domain benchmark with $71{,}237$ trials and $11$ unseen attacks; \textbf{2) test of ASVspoof 2021 LA (\testasvii)}~\cite{liu2023asvspoof2021}, containing $181{,}113$ trials with unseen telephony and codec channel variations; \textbf{3) In-the-Wild (\testitw)}~\cite{muller2022ood}, comprising $31{,}779$ real-world utterances collected from uncurated platforms like YouTube; and \textbf{4) MLAAD v9 (\testml)}~\cite{muller2024mlaad}, a massive multilingual benchmark with approximately $298{,}000$ utterances across $51$ languages and $140$ TTS models.

\noindent\textbf{Evaluation Metrics:}
We report Accuracy and Macro F1-score (hereafter F1) to maintain consistency with the generative classification paradigm of LLMs, where the LLM produces discrete text tokens (\texttt{real} or \texttt{fake}).

\section{Experiment Part I: Audio Encoder and LLM Configuration}
\label{sec:exp:rq1}

The first experiment addresses RQ1 and aims to find the best choice of audio encoder and its integration with the LLM.
We use the LLM for binary detection (real or fake) only; openSMILE features (RQ2) are not investigated in this section.

\begin{table}[!t]
\centering
\caption{Evaluation of audio encoders across integration regimes using a Qwen-0.5B LLM backbone. Individual benchmark columns (LA19, ITW, LA21, ML) report F1-scores (\%), while the last column reports the overall averagd Accuracy / F1-score (\%).
}
\label{tab:encoder_choice}
\setlength{\tabcolsep}{3pt}
\renewcommand{\arraystretch}{1.1}
\resizebox{\columnwidth}{!}{%
\begin{tabular}{@{}lll cccc c@{}}
\toprule
\textbf{Audio Encoder} & \textbf{} & \textbf{LLM} & \testasvi & \testitw & \testasvii & \testml & \textbf{Avg (Acc/F1)} \\
\midrule
\rowcolor{gray!10} \multicolumn{8}{l}{\textit{Discriminative / Self-Supervised Encoders}} \\
\multirow{5}{*}{Whisper}
  & FR      & LoRA-16 & 97.36 & 74.18 & 88.39 & 88.67 & 89.91 / 87.15 \\
\cmidrule(lr){2-8}
  & WA      & Frozen  & 98.41 & 33.52 & 71.42 & 55.97 & 73.61 / 64.83 \\
  & WA      & LoRA-16 & 99.21 & 28.05 & 73.61 & 49.58 & 72.26 / 62.61 \\
\cmidrule(lr){2-8}
  & FT      & Frozen  & 98.84 & 77.65 & 97.39 & 90.50 & \textbf{91.81 / 91.09} \\
  & FT      & LoRA-16 & 99.01 & 81.49 & 97.24 & 68.18 & 87.77 / 86.48 \\
\midrule
\multirow{5}{*}{Wav2Vec2}
  & FR      & LoRA-16 & 96.84 & 33.15 & 89.85 & 56.21 & 74.62 / 69.01 \\
\cmidrule(lr){2-8}
  & WA      & Frozen  & 97.10 & 66.93 & 91.26 & 70.22 & 83.47 / 81.38 \\
  & WA      & LoRA-16 & 87.58 & 72.20 & 87.93 & 70.73 & 83.17 / 79.61 \\
\cmidrule(lr){2-8}
  & FT      & Frozen  & 95.99 & 80.75 & 95.68 & 81.31 & \textbf{89.75 / 88.43} \\
  & FT      & LoRA-16 & 90.88 & 76.19 & 92.71 & 73.98 & 85.90 / 83.44 \\
\midrule
\multirow{3}{*}{EAT}
  & FR      & LoRA-16 & 85.05 & 27.14 & 65.11 & 57.62 & \textbf{70.76 / 58.73} \\
\cmidrule(lr){2-8}
  & FT      & Frozen  & 81.79 & 27.12 & 62.59 & 57.62 & 69.95 / 57.28 \\
  & FT      & LoRA-16 & 75.68 & 27.12 & 72.01 & 56.81 & 67.67 / 57.91 \\
\midrule
\rowcolor{gray!10} \multicolumn{8}{l}{\textit{Neural Audio Codecs}} \\
SpeechTok.$^{\dagger}$ & FR & LoRA-16 & 71.57 & 70.22 & 71.67 & 33.68 & \textbf{71.61 / 61.78} \\
DAC$^{\dagger}$         & FR & LoRA-16 & 68.72 & 52.87 & 67.92 & 38.70 & \textbf{65.90 / 57.05} \\
\bottomrule
\multicolumn{8}{l}{\scriptsize \textbf{FR}: Frozen; \textbf{WA}: Weighted Aggregation; \textbf{FT}: Fine-tuned.} \\
\multicolumn{8}{l}{\scriptsize $^{\dagger}$Uses continuous embeddings from the first $N$ codebooks (\textit{cont:N}). SpeechTokenizer: $N=4$; DAC: $N=6$.}
\vspace{-0.5cm}
\end{tabular}}
\end{table}

\subsection{Model Settings and Training Details}

We employ the Qwen2.5-Instruct LLM family~\cite{qwen2} across three scales ($0.5$B, $3$B, and $7$B) to assess capacity-efficiency trade-offs. Their strong instruction-tuning ensures reliable structured binary outputs.
The evaluated audio encoders include Whisper~\cite{radford2023whisper}, Wav2Vec2~\cite{baevski2020wav2vec}, EAT~\cite{eat}, SpeechTokenizer~\cite{zhang2023speechtokenizer}, and DAC~\cite{kumar2023dac}. 
The LLM is either kept frozen or adapted via LoRA~\cite{hu2022LoRA} with a fixed rank of $16$ and varying alpha values $\alpha \in \{16, 64, 256\}$. The default setting of LoRA, denoted as LoRA-16, utilizes rank $16$ and $\alpha = 64$.

For experiments with fine-tuned audio encoders, the encoder learning rate is $1 \times 10^{-6}$, while LLM LoRA parameters and the projector are optimized with $1 \times 10^{-5}$.
All models are trained for $5$ epochs with a cumulative batch size of $16$ using NVIDIA H100 GPUs.

\subsection{Results on Audio Encoders}
\label{sec:results:rq1}

Table~\ref{tab:encoder_choice} reports evaluation results for the five audio encoder families, comparing three encoder integration strategies (Frozen: FR, Weighted Average: WA, Finetuning: FT) and two LLM adaptation configurations (Frozen, LoRA-16), using Qwen-0.5B as the backbone. We summarize
the observations related to our research question below. 

\textbf{Audio Encoder fine-tuning is the primary driver of out-of-domain generalization.}
Comparing the three encoder integration regimes (FR, WA, FT) in Table~\ref{tab:encoder_choice}, it is evident that relying on frozen (FR) audio encoder representations causes significant degradation on out-of-domain data. For instance, under the LoRA-16 configuration, a frozen Wav2Vec2 encoder achieves near-perfect in-domain performance (\testasvi{} F1: 99.44\%) but collapses on \testitw{} (54.95\%) and drops significantly on \testml{} (69.78\%). Applying end-to-end fine-tuning (FT) to Wav2Vec2 while using the same LoRA-16 setting on LLM (Wav2Vec2 FT+LoRA-16 in Table~\ref{tab:encoder_choice}) improves its \testitw{} F1-score to \textbf{75.60\%} and \testml{} F1-score to \textbf{74.24\%}. Whisper follows the same trend,
where fine-tuning increases the ITW F1-score from 65.73\% to 73.90\% under LoRA-16, reaching a peak of 76.46\% when paired with a frozen LLM.


\textbf{Weighted aggregation is insufficient for LLM-based detectors.}
When examining the WA rows for Whisper with a frozen LLM backbone, weighted aggregation yields a \testitw{} F1-score of only 55.00\%, lagging far behind the fully fine-tuned Whisper (76.46\%). 
Simply re-weighting pre-trained features is not as effective as fine-tuning the audio encoder.

\textbf{Other Audio Encoders underperform Whisper and Wav2Vec2.}
EAT's pretraining on environmental sounds does not transfer effectively to deepfake voice detection.
Its peak out-of-domain F1-scores, achieved under the FT+Frozen regime (\testitw{}: 54.20\%; \testml{}: 69.91\%), 
lag far behind Whisper and Wav2Vec2 under the same configuration 
(Whisper FT+Frozen: \testitw{} 76.46\%, \testml{} 80.45\%; Wav2Vec2 FT+Frozen: \testitw{} 74.79\%, \testml{} 78.98\%), highlighting the critical role of domain alignment in pretraining data.

Under the frozen-encoder, LoRA-16 regime, the neural codecs (SpeechTokenizer and DAC) exhibit the poorest out-of-domain generalization, 
with \testml{} F1-scores of 65.83\% and 65.59\% respectively (Table~\ref{tab:encoder_choice}), significantly trailing other audio encoders. A pontial reason is that the neural audio codecs are inherently optimized for human perceptual transparency, and their vector quantization may discard inaudible artifacts useful for deepfake voice detection.


\subsection{Results on LLM Backbone and Fine-Tuning}
\label{sec:results:scale}

Results in Table~\ref{tab:encoder_choice} suggest that \textbf{a frozen LLM backbone is highly competitive for binary deepfake voice detection when the audio encoder is fine-tuned.}
Holding the LLM entirely \emph{frozen} yields the highest average performance for both Whisper (91.81\% Acc / 91.09\% F1) and Wav2Vec2 (89.75\% Acc / 88.43\% F1), \textbf{provided their respective audio encoders are fine-tuned.}

Interestingly, updating the LLM via LoRA does not consistently improve out-of-domain performance. On \testml{}, comparing the Whisper FT+Frozen and Whisper FT+LoRA-16 rows in Table~\ref{tab:encoder_choice}, LoRA adaptation degrades \testml{} F1 from \textbf{90.50\%} to \textbf{68.18\%}. 
This challenges the prevailing assumption in recent ALLM studies that the LLM must be fine-tuned for deepfake detection when the LLM is prompted with the latent features from fine-tuned audio encoders.

\begin{table}[!t]
\centering
\caption{LLM Scale and LoRA Ranks Impact on Average Performance (Accuracy / F1 \%).}
\label{tab:lora_scaling_summary}
\setlength{\tabcolsep}{4pt}
\renewcommand{\arraystretch}{1.05}
\scalebox{0.9}{
\begin{tabular}{ll ccc}
\toprule
\textbf{Audio Encoder} & \textbf{Config} & \textbf{0.5B} & \textbf{3B} & \textbf{7B} \\
\midrule
\multirow{4}{*}{Wav2Vec2} 
  & Frozen   & \textbf{89.75 / 88.43} & \textbf{89.64 / 87.64} & \textbf{89.77 / 88.63} \\
  & LoRA-16  & 85.90 / 83.44 & 86.38 / 84.23 & 86.11 / 84.44 \\
  & LoRA-64  & 86.67 / 83.81 & 88.93 / 87.62 & 85.33 / 84.52 \\
  & LoRA-256 & 83.08 / 81.28 & 86.59 / 85.25 & 85.39 / 83.05 \\
\midrule
\multirow{4}{*}{Whisper}  
  & Frozen   & \textbf{91.81 / 91.09} & \textbf{92.47 / 90.94} & \textbf{91.59 / 90.58} \\
  & LoRA-16  & 87.77 / 86.48 & 91.48 / 88.75 & 90.36 / 88.67 \\
  & LoRA-64  & 88.36 / 85.85 & 87.85 / 87.14 & 90.01 / 88.64 \\
  & LoRA-256 & 89.49 / 86.88 & 88.97 / 87.47 & 89.82 / 88.42 \\
\bottomrule
\end{tabular}}
\vspace{-0.4cm}
\end{table}

To verify that this finding is not due to a specific LoRA configuration or LLM size, we systematically varied the LoRA alpha ($\alpha \in \{16, 64, 256\}$) across multiple LLM backbones (0.5B, 3B, 7B). As summarized in Table~\ref{tab:lora_scaling_summary}, 
a frozen LLM consistently outperformed its counterparts that are tuned with LoRA across all evaluated configurations.  

These results also demonstrate that \textbf{scaling the LLM backbone does not improve detection performance when the LLM remains frozen.}
While increasing model capacity is typically assumed to be beneficial, scaling the frozen LLM from 0.5B to 3B and 7B yields no consistent improvements. For instance, the Average F1 for the Whisper FT + Frozen configuration slightly degrades from 91.09\% at 0.5B to 90.94\% at 3B and 90.58\% at 7B. A similarly flat or fluctuating trend is observed with Wav2Vec2, confirming that brute-force scaling is not a silver bullet for out-of-domain generalization.

In summary for RQ1, Wav2Vec2 and Whisper are better choices than EAT and neural codecs as the audio encoder. Interestingly, contrary to the prevailing consensus in standalone deepfake detection, where Wav2Vec2 is typically the dominant audio encoder, our results show that a fine-tuned Whisper audio encoder is highly effective and often superior in the ALLM context. This suggests that Whisper's linguistically aligned representations form a stronger synergy with the LLM's semantic reasoning space. Additionally, fine-tuning the audio encoder improves out-of-domain detection performance while maintaining in-domain performance. In contrast, fine-tuning of the LLM alone
is insufficient and risks out-of-domain degradation. A frozen instruct-LLM paired with a fine-tuned SSL encoder offers a strong and resource-efficient alternative. 

This limitation naturally motivates our next experiment: whether explicitly bridging the modality gap via text-based acoustic features can unlock the potential of LLM adaptation.



\begin{table}[htbp]
\centering
\caption{Effect of adding openSMILE (OS) features.}
\label{tab:result:q2}
\setlength{\tabcolsep}{2.5pt}
\renewcommand{\arraystretch}{1.15}
\resizebox{\columnwidth}{!}{%
\begin{tabular}{@{}ll c cccc c@{}}
\toprule
\textbf{Enc.} & \textbf{LLM} & \textbf{OS} & \testasvi & \testitw & \testasvii & \testml & \textbf{Avg (Acc / F1)} \\
\midrule
\rowcolor{gray!15} \multicolumn{8}{c}{\textit{Frozen LLM}} \\
\multirow{6}{*}{\rotatebox{90}{\shortstack[l]{Wav2Vec2 \\ \scriptsize (FT)}}}
  & \multirow{2}{*}{0.5B} & $\times$ & 95.99 & 80.75 & 95.68 & 81.31 & 89.75 / 88.43 \\
  & & $\checkmark$ & 99.64 & 77.27 & 96.82 & 91.84 & 91.98 (\textbf{+2.23}) / 91.39 (\textbf{+2.96}) \\
\cmidrule(lr){2-8}
  & \multirow{2}{*}{3B} & $\times$ & 93.02 & 84.16 & 94.37 & 79.00 & 89.64 / 87.64 \\
  & & $\checkmark$ & 97.87 & 84.36 & 96.16 & 83.45 & 91.46 (\textbf{+1.82}) / 90.46 (\textbf{+2.82}) \\
\cmidrule(lr){2-8}
  & \multirow{2}{*}{7B} & $\times$ & 96.59 & 84.16 & 96.77 & 76.99 & 89.77 / 88.63 \\
  & & $\checkmark$ & 99.28 & 84.36 & 94.45 & 70.26 & 88.42 ($-$1.35) / 87.09 ($-$1.54) \\
\midrule
\multirow{6}{*}{\rotatebox{90}{\shortstack[l]{Whisper \\ \scriptsize (FT)}}}
  & \multirow{2}{*}{0.5B} & $\times$ & 98.84 & 77.65 & 97.39 & 90.50 & 91.81 / 91.09 \\
  & & $\checkmark$ & 99.12 & 83.96 & 97.67 & 93.42 & 94.15 (\textbf{+2.34}) / 93.54 (\textbf{+2.45}) \\
\cmidrule(lr){2-8}
  & \multirow{2}{*}{3B} & $\times$ & 97.67 & 83.15 & 96.22 & 86.71 & 92.47 / 90.94 \\
  & & $\checkmark$ & 99.08 & 81.25 & 95.34 & 91.91 & 93.42 (\textbf{+0.95}) / 91.89 (\textbf{+0.95}) \\
\cmidrule(lr){2-8}
  & \multirow{2}{*}{7B} & $\times$ & 98.40 & 84.00 & 96.01 & 83.91 & 91.59 / 90.58 \\
  & & $\checkmark$ & 98.34 & 81.01 & 96.36 & 91.85 & 92.79 (\textbf{+1.20}) / 91.89 (\textbf{+1.31}) \\
\midrule
\rowcolor{gray!15} \multicolumn{8}{c}{\textit{LoRA-16}} \\
\multirow{6}{*}{\rotatebox{90}{\shortstack[l]{Wav2Vec2 \\ \scriptsize (FT)}}}
  & \multirow{2}{*}{0.5B} & $\times$ & 90.88 & 76.19 & 92.71 & 73.98 & 85.90 / 83.44 \\
  & & $\checkmark$ & 96.42 & 74.51 & 96.42 & 76.36 & 87.04 (\textbf{+1.14}) / 85.93 (\textbf{+2.49}) \\
\cmidrule(lr){2-8}
  & \multirow{2}{*}{3B} & $\times$ & 93.35 & 77.29 & 92.71 & 73.58 & 86.38 / 84.23 \\
  & & $\checkmark$ & 97.78 & 70.14 & 96.83 & 83.96 & 88.05 (\textbf{+1.67}) / 87.18 (\textbf{+2.95}) \\
\cmidrule(lr){2-8}
  & \multirow{2}{*}{7B} & $\times$ & 94.33 & 74.97 & 94.83 & 73.62 & 86.11 / 84.44 \\
  & & $\checkmark$ & 97.06 & 82.62 & 97.34 & 81.69 & 90.64 (\textbf{+4.53}) / 89.68 (\textbf{+5.24}) \\
\midrule
\multirow{6}{*}{\rotatebox{90}{\shortstack[l]{Whisper \\ \scriptsize (FT)}}}
  & \multirow{2}{*}{0.5B} & $\times$ & 99.01 & 81.49 & 97.24 & 68.18 & 87.77 / 86.48 \\
  & & $\checkmark$ & 99.47 & 87.53 & 96.96 & 91.17 & 94.49 (\textbf{+6.72}) / 93.78 (\textbf{+7.30}) \\
\cmidrule(lr){2-8}
  & \multirow{2}{*}{3B} & $\times$ & 89.71 & 83.96 & 93.61 & 87.74 & 91.48 / 88.75 \\
  & & $\checkmark$ & 98.53 & 87.67 & 96.39 & 80.19 & 91.68 (\textbf{+0.20}) / 90.69 (\textbf{+1.94}) \\
\cmidrule(lr){2-8}
  & \multirow{2}{*}{7B} & $\times$ & 95.22 & 86.53 & 95.41 & 77.54 & 90.36 / 88.67 \\
  & & $\checkmark$ & 98.80 & 86.16 & 93.94 & 84.55 & 92.17 (\textbf{+1.81}) / 90.86 (\textbf{+2.19}) \\
\midrule
\rowcolor{gray!15} \multicolumn{8}{c}{\textit{End-to-End Multimodal}} \\
\multirow{2}{*}{\rotatebox{90}{\shortstack[l]{Omni \\ \scriptsize (FT)}}} & \multirow{2}{*}{3B}
  & $\times$ & 91.71 & 28.54 & 77.46 & 48.47 & 80.50 / 61.54 \\
  & & $\checkmark$ & 95.90 & 27.73 & 78.68 & 48.97 & 81.48 (\textbf{+0.98}) / 62.82 (\textbf{+1.28}) \\
\bottomrule
\end{tabular}}
\vspace{-15pt}
\end{table}

\section{Experiment Part II: Effectiveness of openSMILE Feature Injection}
\label{sec:exp:rq2}

To address RQ2, we investigate whether prompting with textual description of the acoustic features via openSMILE bridges the modality gap.

\subsection{Model Settings and Training Details}

We evaluate four base configurations pairing two audio encoders (fine-tuned Whisper, fine-tuned Wav2Vec2) with the two LLM settings identified as most informative in Experiment I: the frozen LLM and LoRA-16. We test at three LLM scales (0.5B, 3B, and 7B) and include Qwen2.5-Omni-3B~\cite{qwenomni} as an end-to-end reference. For Qwen2.5-Omni-3B, we implement LoRA adaptation following the protocol in~\cite{xie2026ftgrpo} (rank $16$ and $\alpha = 64$). Each configuration is compared with and without the openSMILE text prompt. 

\subsection{Results}
\label{sec:results:rq2}

Table~\ref{tab:result:q2} details the impact of openSMILE-based prompt across encoder families and LLM adaptation strategies. We summarize the observations below.

\textbf{Injecting explicit openSMILE features into a frozen LLM yields good out-of-domain robustness with zero LLM-updating costs.} For instance, appending openSMILE to the Whisper-backed Qwen-0.5B (Frozen) model improves Average F1 by \textbf{2.45\%} (\textbf{91.09\% $\to$ 93.54\%}) and boosts performance on the \testml{} benchmark by \textbf{2.92\%} (\textbf{90.50\% $\to$ 93.42\%}). 

\textbf{Prompt based on openSMILE acts as a cross-modal alignment bridge.}
Although LoRA adaptation was found to be ineffective in Experiment I, when explicit openSMILE text descriptors are injected, LoRA adaptation becomes highly synergistic. {For instance, using the 0.5B Whisper-backed model under LoRA-16 adaptation}, the \testml{} F1 improves by \textbf{22.99\%} (\textbf{68.18\% $\to$ 91.17\%}), fully recovering and surpassing the no-openSMILE frozen baseline. Because the LLM is provided with openSMILE acoustic features in its native textual modality, we assume that the LoRA parameters no longer merely memorize training artifacts, but instead act as a cross-modal alignment bridge.

\textbf{Observations on native multimodal integration.}
For the natively integrated Qwen2.5-Omni-3B model, openSMILE injection yields marginal improvements \textbf{(e.g., +1.28\% in Average F1)}; this might be attributed to the lower temporal resolution of its integrated audio encoder (12.5 Hz) compared to that of the encoders used in our configurations (50 Hz).

\textbf{Audio Encoder pretraining dictates the impact of explicit features.}
The magnitude of performance gains from openSMILE injection is directly tied to the audio encoder's pretraining objective. Whisper, optimized for ASR, inherently suppresses non-linguistic micro-artifacts. Consequently, explicit openSMILE text tokens provide critical missing forensic evidence, yielding transformative improvements: the 0.5B Whisper-backed model under LoRA-16 achieves an Average F1 of \textbf{93.78\%} (up from \textbf{86.48\%}, a gain of \textbf{7.30\%}). Conversely, Wav2Vec2's self-supervised pretraining natively preserves low-level acoustic variation directly from the raw waveform. Therefore, explicit openSMILE prompts yield modest, complementary gains (e.g., \textbf{+2.49\%} Average F1 for the 0.5B Wav2Vec2 model under LoRA-16 adaptation).

In summary for RQ2, openSMILE features improve the generalization performance of the LLM-based detector. They also mitigate the degradation caused by fine-tuning the LLM backbone, when compared with the case of using a frozen LLM.

\section{Comparison with State of the Art Systems}
\label{sec:compare}

Table~\ref{tab:sota} compares our approach against several state-of-the-art classical, SSL-based, and ALLM-based systems. We report the results on the unseen out-of-domain data: \testitw{} and \testml{}. This is because all evaluated ALLM-based systems achieved F1-scores approaching 100\% on the in-domain \testasvi{} dataset, but their performance varies considerably on these out-of-domain subsets. We report Macro-F1 (\%) to maintain consistency with other reported results. Our observations are summarized below.

\begin{table}[htbp]
\centering
\caption{Comparison with SOTA systems on unseen domains (F1,\%).}
\label{tab:sota}
\setlength{\tabcolsep}{3pt}
\resizebox{\columnwidth}{!}{%
\begin{tabular}{@{}ll cc@{}}
\toprule
\textbf{Method} & \textbf{\#Params} & \textbf{\testitw} & \textbf{\testml} \\
\midrule
\rowcolor{gray!10} \multicolumn{4}{l}{\textit{Classical \& SSL-based Prior Work}} \\
Whisper FT & 0.6B & 71.11 & 67.49  \\
Wav2Vec2-Conformer~\cite{truong2024temporal} & 0.3B & 64.40 & 71.60 \\
Wav2Vec2-Mamba~\cite{xiao2025xlsr}            & 0.3B & 52.30 & 55.80 \\
Wav2Vec2-AASIST~\cite{tak2022wav2vec}        & 0.3B & 69.10 & 79.80 \\
\midrule
\rowcolor{gray!10} \multicolumn{4}{l}{\textit{ALLM-based and Retrieval Prior Work}} \\
ALLM4ADD~\cite{gu2025allm4add}           & 7B  & 68.34 & 77.87 \\
FT-GRPO~\cite{xie2026ftgrpo}             & 3B  & 28.54 & 66.14 \\
ICLAD~\cite{chou2026iclad}               & N/A.$^\dagger$ & 77.80 & 59.30 \\
\midrule
\rowcolor{blue!5} \multicolumn{4}{l}{\textit{Ours (Explicit Textual Grounding)}} \\
Wav2Vec2 FT + Qwen-0.5B Frozen + OS & 0.9B & 77.27 & 91.84 \\
Whisper FT + Qwen-0.5B Frozen + OS  & 1.1B & 83.96 & \textbf{93.42} \\
Whisper FT + Qwen-0.5B LoRA-16 + OS & 1.1B & \textbf{87.53} & 91.17 \\
\bottomrule
\multicolumn{4}{l}{\scriptsize $^\dagger$ Gemini-2.5 Flash is a proprietary model; exact parameters undisclosed.}
\end{tabular}}
\vspace{-0.6cm}
\end{table}

\textbf{Enhancing SSL Encoders with a Frozen LLM.} 
Comparison between our Wav2Vec2-based configuration and classical SSL-based prior work highlights the direct benefit of integrating an LLM. While popular architectures utilizing similar SSL encoders and traditional classifiers (such as W2V2-AASIST) achieve up to 79.80\% F1 on \testml{}, replacing these classifiers with our textually grounded LLM yields superior generalization. Our \textit{Wav2Vec2 FT + Qwen-0.5B Frozen + OS} configuration pushes performance further, reaching 77.27\% on \testitw{} and 91.84\% on \testml{}. This demonstrates that appending a lightweight frozen LLM backbone to an acoustic encoder is highly beneficial.

\textbf{Comparison with Standalone and ALLM Strategies.} 
The benefit of explicit openSMILE-based textual prompt becomes evident when comparing our Whisper-based models against both standalone encoders and recent ALLM baselines. Relying solely on a fine-tuned Whisper encoder (\textit{Whisper FT}) yields limited out-of-domain generalization, achieving only 71.11\% on \testitw{} and 67.49\% on \testml{}. However, integrating this exact encoder with a lightweight LLM and openSMILE descriptors triggers a massive jump: our frozen configuration achieves \textbf{83.96\%} on \testitw{} and \textbf{93.42\%} on \testml{}. 

When comparing our Whisper-based models with recent ALLM baselines, as shown in Table~\ref{tab:sota}, our framework of injecting explicit openSMILE-based acoustic descriptions achieved competitive generalization, reaching an \textbf{87.53\%} F1-score on \testitw{} and \textbf{91.17\%} on \testml{} (when utilizing LoRA-16). Furthermore, our framework remains highly efficient, with lightweight configurations totaling approximately 1.1B parameters, demonstrating that our approach is an effective alternative to larger or more complex ALLM strategies.

\vspace{-0.1cm}
\section{Conclusion}
\label{sec:conclusion}

We presented an ALLM framework for deepfake voice detection that bridges the modality gap between acoustic physics and LLM semantics. Our evaluation demonstrates that optimizing the audio encoder allows a frozen LLM to generalize robustly without overfitting. Furthermore, injecting openSMILE acoustic descriptors as structured text tokens provides crucial semantic context. This cross-modal grounding enhances the frozen baseline and synergizes with LoRA adaptation, balancing computational efficiency with peak performance. Ultimately, this framework paves the way for transparent, reasoning-based audio forensics, where interpretable acoustic statistics enable explainable detection. Future work will investigate prompt design and advanced alternatives for cross-modal alignment.

\newpage


\bibliographystyle{IEEEtran}
\bibliography{custom}

@inproceedings{gu2025allm4add,
  title={{ALLM4ADD}: Unlocking the Capabilities of Audio Large Language Models for Audio Deepfake Detection},
  author={Gu, Hao and Yi, Jiangyan and Wang, Chenglong and Tao, Jianhua and Lian, Zheng and He, Jiayi and Ren, Yong and Chen, Yujie and Wen, Zhengqi},
  booktitle={Proceedings of the 33rd ACM International Conference on Multimedia},
  pages={11736--11745},
  year={2025}
}

@article{li2025dfallm,
  title={{DFALLM}: Achieving Generalizable Multitask Deepfake Detection by Optimizing Audio {LLM} Components},
  author={Li, Yupei and Wang, Li and Wang, Yuxiang and Wang, Lei and Cai, Rizhao and Shi, Jie and Schuller, Bj{\"o}rn and Wu, Zhizheng},
  journal={arXiv preprint arXiv:2512.08403},
  year={2025}
}

@article{xie2026ftgrpo,
  title={Interpretable All-Type Audio Deepfake Detection with Audio {LLMs} via Frequency--Time Reinforcement Learning},
  author={Xie, Yuankun and Guo, Xiaoxuan and Zhou, Jiayi and Wang, Tao and Liu, Jian and Fu, Ruibo and Wang, Xiaopeng and Cheng, Haonan and Ye, Long},
  journal={arXiv preprint arXiv:2601.02983},
  year={2026}
}

@article{guo2026sddapallm,
  title={Towards Explicit Acoustic Evidence Perception in Audio {LLMs} for Speech Deepfake Detection},
  author={Guo, Xiaoxuan and Xie, Yuankun and Cheng, Haonan and Zhou, Jiayi and Liu, Jian and Huang, Hengyan and Ye, Long and Zhang, Qin},
  journal={arXiv preprint arXiv:2601.23066},
  year={2026}
}

@inproceedings{jung2022aasist,
  title={{AASIST}: Audio Anti-Spoofing Using Integrated Spectro-Temporal Graph Attention Networks},
  author={Jung, Jee-weon and Heo, Hee-Soo and Tak, Hemlata and Shim, Hye-jin and Chung, Joon Son and Lee, Bong-Jin and Yu, Ha-Jin and Evans, Nicholas},
  booktitle={Proc. ICASSP 2022},
  pages={6367--6371},
  year={2022}
}

@inproceedings{tak2022wav2vec,
  title={Automatic Speaker Verification Spoofing and Deepfake Detection using {wav2vec 2.0} and Data Augmentation},
  author={Tak, Hemlata and Todisco, Massimiliano and Wang, Xin and Jung, Jee-weon and Yamagishi, Junichi and Evans, Nicholas},
  year      = {2022},
  booktitle = {Proc. Odyssey 2022},
  pages     = {112--119},
}

@inproceedings{ogunremiTranscribe2025,
	address = {Honolulu, HI, USA},
	title = {Transcribe, {Translate}, or {Transliterate}: {An} {Investigation} of {Intermediate} {Representations} in {Spoken} {Language} {Models}},
	copyright = {https://doi.org/10.15223/policy-029},
	isbn = {979-8-3315-4426-3},
	shorttitle = {Transcribe, {Translate}, or {Transliterate}},
	url = {https://ieeexplore.ieee.org/document/11434765/},
	doi = {10.1109/ASRU65441.2025.11434765},
	urldate = {2026-06-17},
	booktitle = {Proc. ASRU},
	author = {Ògúnremí, Tolúlọpé and Manning, Christopher D. and Jurafsky, Dan and Livescu, Karen},
	month = dec,
	year = {2025},
	pages = {1--7},
}

@inproceedings{kheirDeepFense2026,
	title = {{DeepFense}: {A} {Unified}, {Modular}, and {Extensible} {Framework} for {Robust} {Deepfake} {Audio} {Detection}},
	url = {https://arxiv.org/abs/2604.08450},
	booktitle = {Proc. {Interspeech}},
	author = {Kheir, Yassine El and Das, Arnab and Xiao, Yixuan and Wang, Xin and Kallel, Feidi and Erdogan, Enes Erdem and Vu, Ngoc Thang and Polzehl, Tim and Moeller, Sebastian},
	year = {2026},
    pages = {(accepted)}
}

@inproceedings{chung_w2v-bert_2021,
	address = {Cartagena, Colombia},
	title = {w2v-{BERT}: {Combining} {Contrastive} {Learning} and {Masked} {Language} {Modeling} for {Self}-{Supervised} {Speech} {Pre}-{Training}},
	copyright = {https://doi.org/10.15223/policy-029},
	isbn = {978-1-6654-3739-4},
	shorttitle = {w2v-{BERT}},
	url = {https://ieeexplore.ieee.org/document/9688253/},
	doi = {10.1109/ASRU51503.2021.9688253},
	urldate = {2026-05-20},
	booktitle = {Proc. ASRU},
	publisher = {IEEE},
	author = {Chung, Yu-An and Zhang, Yu and Han, Wei and Chiu, Chung-Cheng and Qin, James and Pang, Ruoming and Wu, Yonghui},
	month = dec,
	year = {2021},
	pages = {244--250},
}

@inproceedings{pascu_easy_2025,
	title = {Easy, {Interpretable}, {Effective}: {openSMILE} for voice deepfake detection},
	booktitle = {Proc. {ICASSP} 2025},
	author = {Pascu, Octavian and Oneaţă, Dan and Cucu, Horia and Müller, Nicolas M.},
	year = {2025},
	pages = {1--5},
}

@misc{xuHoliAntiSpoof2026,
	title = {{HoliAntiSpoof}: {Audio} {LLM} for {Holistic} {Speech} {Anti}-{Spoofing}},
	author = {Xu, Xuenan and Ren, Yiming and Liu, Liwei and Wu, Wen and Li, Baoxiang and Lu, Chaochao and Wang, Shuai and Zhang, Chao},
	year = {2026},
}

@article{liSurvey2025,
	title = {A {Survey} on {Speech} {Deepfake} {Detection}},
	volume = {57},
	language = {en},
	number = {7},
	journal = {ACM Computing Surveys},
	publisher = {Association for Computing Machinery (ACM)},
	author = {Li, Menglu and Ahmadiadli, Yasaman and Zhang, Xiao-Ping},
	year = {2025},
	pages = {1--38},
}

@book{tanNeural2023,
	address = {Singapore},
	series = {Artificial {Intelligence}: {Foundations}, {Theory}, and {Algorithms}},
	title = {Neural {Text}-to-{Speech} {Synthesis}},
	copyright = {https://www.springernature.com/gp/researchers/text-and-data-mining},
	isbn = {978-981-99-0826-4 978-981-99-0827-1},
	url = {https://link.springer.com/10.1007/978-981-99-0827-1},
	doi = {10.1007/978-981-99-0827-1},
	language = {en},
	urldate = {2024-05-16},
	publisher = {Springer Nature Singapore},
	author = {Tan, Xu},
	year = {2023},
}

@article{chenNeural2025,
	title = {Neural {Codec} {Language} {Models} are {Zero}-{Shot} {Text} to {Speech} {Synthesizers}},
	volume = {33},
	copyright = {https://ieeexplore.ieee.org/Xplorehelp/downloads/license-information/IEEE.html},
	issn = {2998-4173},
	url = {https://ieeexplore.ieee.org/document/10842513/},
	doi = {10.1109/TASLPRO.2025.3530270},
	urldate = {2025-04-08},
	journal = {IEEE Transactions on Audio, Speech and Language Processing},
	author = {Chen, Sanyuan and Wang, Chengyi and Wu, Yu and Zhang, Ziqiang and Zhou, Long and Liu, Shujie and Chen, Zhuo and Liu, Yanqing and Wang, Huaming and Li, Jinyu and He, Lei and Zhao, Sheng and Wei, Furu},
	year = {2025},
	pages = {705--718},
}

@inproceedings{radford2023whisper,
  title={Robust Speech Recognition via Large-Scale Weak Supervision},
  author={Radford, Alec and Kim, Jong Wook and Xu, Tao and Brockman, Greg and McLeavey, Christine and Sutskever, Ilya},
  booktitle = {Proceedings of the 40th International Conference on Machine Learning},
  numpages = {27},
  year={2023}
}

@inproceedings{baevski2020wav2vec,
  title={{wav2vec 2.0}: A Framework for Self-Supervised Learning of Speech Representations},
  author={Baevski, Alexei and Zhou, Yuhao and Mohamed, Abdelrahman and Auli, Michael},
  booktitle={Advances in Neural Information Processing Systems},
  volume={33},
  pages={12449--12460},
  year={2020}
}

@inproceedings{todisco2019asvspoof,
  title={{ASVspoof 2019}: Future Horizons in Spoofed and Fake Audio Detection},
  author={Todisco, Massimiliano and Wang, Xin and Vestman, Ville and Sahidullah, Md and Delgado, H{\'e}ctor and Nautsch, Andreas and Yamagishi, Junichi and Evans, Nicholas and Kinnunen, Tomi H and Lee, Kong Aik},
  booktitle={Proc. Interspeech 2019},
  pages={1008--1012},
  year={2019}
}

@article{liu2023asvspoof2021,
  title={{ASVspoof 2021}: Towards Spoofed and Deepfake Speech Detection in the Wild},
  author={Liu, Xuechen and Wang, Xin and Sahidullah, Md and Patino, Jose and Delgado, H{\'e}ctor and Kinnunen, Tomi and Todisco, Massimiliano and Yamagishi, Junichi and Evans, Nicholas and Nautsch, Andreas and others},
  journal={IEEE/ACM Transactions on Audio, Speech, and Language Processing},
  year={2023}
}

@inproceedings{wang2024asvspoof5,
  title={{ASVspoof 5}: Crowdsourced Speech Data, Deepfakes, and Adversarial Attacks at Scale},
  author={Wang, Xin and Delgado, H{\'e}ctor and Tak, Hemlata and Jung, Jee-weon and Shim, Hye-jin and Todisco, Massimiliano and Kukanov, Ivan and Liu, Xuechen and Sahidullah, Md and Kinnunen, Tomi and others},
  booktitle={ASVspoof Workshop 2024},
  year={2024},
  pages     = {1--8},
}

@inproceedings{muller2022ood,
  title={Does Audio Deepfake Detection Generalize?},
  author={M{\"u}ller, Nicolas M. and Czempin, Pavel and Dieckmann, Franziska and Froghyar, Adam and B{\"o}ttinger, Konstantin},
  booktitle = {Proc. {Interspeech 2022}},
  pages     = {2783--2787},
  year={2022}
}

@article{muller2024mlaad,
  title={{MLAAD}: The Multi-Language Audio Anti-Spoofing Dataset},
  author={M{\"u}ller, Nicolas M. and Kawa, Piotr and Choong, Wei Herng and Casanova, Edresson and Golge, Eren and M{\"u}ller, Thorsten and Syga, Piotr and Sperl, Philip and B{\"o}ttinger, Konstantin},
  journal={International Joint Conference on Neural Networks},
  year={2024}
  
}

@inproceedings{eyben2010opensmile,
  title={{openSMILE} -- The {M}unich Versatile and Fast Open-Source Audio Feature Extractor},
  author={Eyben, Florian and W{\"o}llmer, Martin and Schuller, Bj{\"o}rn},
  booktitle={Proceedings of the 18th ACM International Conference on Multimedia},
  pages={1459--1462},
  year={2010}
}

@inproceedings{hu2022lora,
  title={{LoRA}: Low-Rank Adaptation of Large Language Models},
  author={Hu, Edward J. and Shen, Yelong and Wallis, Phillip and Allen-Zhu, Zeyuan and Li, Yuanzhi and Wang, Shean and Wang, Lu and Chen, Weizhu},
  booktitle={International Conference on Learning Representations},
  year={2022},
}

@inproceedings{kumar2023dac,
  title={High-Fidelity Audio Compression with Improved {RVQGAN}},
  author={Kumar, Rithesh and Seetharaman, Prem and Luebs, Alejandro and Kumar, Ishaan and Kumar, Kundan},
  booktitle={Advances in Neural Information Processing Systems},
  year={2023}
}

@inproceedings{zhang2023speechtokenizer,
  title={{SpeechTokenizer}: Unified Speech Tokenizer for Speech Language Models},
  author={Zhang, Xin and Zhang, Dong and Li, Shimin and Zhou, Yaqian and Qiu, Xipeng},
  booktitle={The Twelfth International Conference on Learning Representations},
  year={2024}
}

@article{yang2019lfcc,
  title={Significance of Subband Features for Synthetic Speech Detection},
  author={Yang, Jichen and Das, Rohan Kumar and Li, Haizhou},
  journal={IEEE Transactions on Information Forensics and Security},
  volume={15},
  pages={2160--2170},
  year={2019}
}

@inproceedings{tak2021rawnet2,
  title={End-to-End Anti-Spoofing with {RawNet2}},
  author={Tak, Hemlata and Patino, Jose and Todisco, Massimiliano and Nautsch, Andreas and Evans, Nicholas and Larcher, Anthony},
  booktitle={Proc. ICASSP 2021},
  pages={6369--6373},
  year={2021}
}

@article{truong2024temporal,
  title={Temporal-channel modeling in multi-head self-attention for synthetic speech detection},
  author={Truong, Duc-Tuan and Tao, Ruijie and Nguyen, Tuan and Luong, Hieu-Thi and Lee, Kong Aik and Chng, Eng Siong},
  journal={arXiv preprint arXiv:2406.17376},
  year={2024}
}

@article{xiao2025xlsr,
  title={XLSR-Mamba: A dual-column bidirectional state space model for spoofing attack detection},
  author={Xiao, Yang and Das, Rohan Kumar},
  journal={IEEE Signal Processing Letters},
  year={2025},
  publisher={IEEE}
}

@article{chou2026iclad,
  title={ICLAD: In-Context Learning with Comparison-Guidance for Audio Deepfake Detection},
  author={Chou, Benjamin and Zhu, Yi and Koppisetti, Surya},
  journal={arXiv preprint arXiv:2604.16749},
  year={2026}
}

@article{qwenomni,
  title={Qwen2.5-Omni Technical Report},
  author={QwenTeam},
  journal={arXiv preprint},
  year={2025}
}

@article{qwen2,
      title={Qwen2 Technical Report}, 
      author={An Yang and Baosong Yang and Binyuan Hui and Bo Zheng and Bowen Yu and Chang Zhou and Chengpeng Li and others},
      journal={arXiv preprint arXiv:2407.10671},
      year={2024}
}

@inproceedings{eat,
  title={Self-Supervised Audio Teacher-Student Transformer for Both Clip-Level and Frame-Level Tasks},
  author={Chen, Bin and Luo, Zi-Ang and Wang, Chao and Pan, Zhiqiang and Tian, Weidi},
  booktitle={Proceedings of the 32nd ACM International Conference on Multimedia},
  year={2024}
}

@inproceedings{xiaoLayerWise2025,
  title = {Layer-{{Wise Decision Fusion}} for {{Fake Audio Detection Using XLS-R}}},
  booktitle = {Proc. Interspeech 2025},
  author = {Xiao, Yixuan and Vu, Ngoc Thang},
  pages = {5618--5622},
  year = {2025},
}

@inproceedings{kheir2025,
    title = {{Comprehensive Layer-wise Analysis of {SSL} Models for Audio Deepfake Detection}},
    author = {El Kheir, Yassine  and
      Samih, Younes  and
      Maharjan, Suraj  and
      Polzehl, Tim  and
      M{\"o}ller, Sebastian},
    editor = "Chiruzzo, Luis  and
      Ritter, Alan  and
      Wang, Lu",
    booktitle = "Findings of the Association for Computational Linguistics: NAACL 2025",
    year = "2025",
    pages = "4070--4082",
}

\end{document}